\documentclass[aps,prb,twocolumn,preprintnumbers,amsmath,amssymb,longbibliography]{revtex4-1}

\usepackage{bbm}
\usepackage{bm}
\usepackage{color}
\usepackage{dcolumn}
\usepackage{epstopdf}
\usepackage{graphicx}
\usepackage[colorlinks, citecolor=blue]{hyperref}
\usepackage{lipsum}
\usepackage{float}
\usepackage[export]{adjustbox}

\newcommand{\beq}{ \begin{equation} }
\newcommand{\eeq}{ \end{equation} }

\newcommand{\e}{\varepsilon}

\begin{document}

\title{Current cross-correlations and waiting time distributions in Andreev transport through Cooper pair splitters
based on triple quantum dots}

\author{Kacper Wrze\'sniewski}
\email{wrzesniewski@amu.edu.pl}

\author{Ireneusz Weymann}
\affiliation{Department of Mesoscopic Physics, Faculty of Physics,
	Adam Mickiewicz University, ul.~Uniwersytetu Pozna\'nskiego 2, 61-614 Pozna{\'n}, Poland}

\begin{abstract}
We study the spin-resolved subgap transport in a triple quantum-dot
system coupled to one superconducting and two ferromagnetic
leads. We examine the Andreev processes in the parallel
and antiparallel alignments of ferromagnets magnetic moments
in both the linear and nonlinear response regimes. The emphasis
is put on the analysis of the current cross-correlations
between the currents flowing through the left and
right arms of the device and relevant electron waiting time distributions.
We show that both quantities can give
an important insight into the subgap transport processes and their
analysis can help optimizing the system parameters for achieving the considerable Andreev current and efficient
Cooper pair splitting. Strong positive values
of cross-correlations are associated with the presence of tunneling
processes enhancing the Cooper pair splitting efficiency,
while short waiting times for electrons
tunneling through distinct ferromagnetic contacts indicate
fast splitting of emitted Cooper pairs.
In particular, we study two detuning schemes and show
that antisymmetric shift of side quantum dots energy levels is favorable for efficient Cooper pair splitting.
The analysis of spin-resolved waiting time distributions supports
the performance enhancement due to the presence of ferromagnetic contacts,
which is in particular revealed for short times.
Finally, we consider the effect of changing the inter-dot hopping amplitude
and predict that strong inter-dot correlations lead to a reduction of Andreev transport properties in low-bias limit.
\end{abstract}

\maketitle

\section{Introduction}

Rapid progress in nanofabrication and experimental techniques in condensed matter brings the possibility to prepare and study highly-tunable hybrid structures in a controllable and precise manner.
\cite{DeFranceschi2010,Rodero2011}
One prominent class of hybrid nanoscale devices are quantum dots coupled to superconducting and normal contacts.
\cite{Choi2000,Feng2003,Cao2004,Csonka2010,SiqueiraPRB10,Lee2012,Lee2014,Kumar2014}
In this regard, particularly interesting is a three-terminal setup
known as Cooper pair splitter (CPS),
\cite{DLoss2001, Russo2005, Hofstetter2009, Futterer2009, Herrmann2010, Schindele2012, Schindele2014, Wojcik2014, Gramich2015, Gramich2016}
which consists of one superconducting and two normal metallic leads.
Such CPS devices provide the possibility to generate non-local pairs of entangled electrons in a controllable fashion.
As far as the transport properties and splitting efficiency of a CPS are concerned,
the most promising transport regime of the system is when the applied bias voltage
is smaller than the superconducting energy gap $\Delta$.
In such subgap transport regime, the current flows through the device
by the processes known as Andreev reflections.\cite{Andreev1964, Rodero2011}
In the beam splitter geometry, the two dominant processes
are the direct (DAR) and crossed (CAR) Andreev reflections. \cite{Beckmann2004}
The first type of processes concerns transport of two electrons forming a Cooper pair
through the same arm of the device, whereas the second type describes the transfer of a Cooper pair,
which is split into distinct drain electrodes. For an optimal operation
of a CPS it is therefore crucial to engineer devices, in which
Andreev current flows mainly due to CAR processes.
This can be achieved by appropriate choice of nanostructure's size and geometry
as well as fine tuning by gate voltages.\cite{Hofstetter2009,Schindele2012}

The transport properties of Cooper pair splitters based on double quantum dots have already been widely studied,
both theoretically and experimentally. \cite{SiqueiraPRB10, Eldridge2010, Hoffstetter2011, Schindele2012, Schindele2014, Sothmann2014, Fulop2014, Trocha2014, Trocha2015, Tan2015, Hussein2016, Su2017, Wrzesniewski2017cc, Wrzesniewski2017kondo}
High splitting efficiency in these models is obtained due to
a high difference between inter and intra-dot Coulomb interactions
and through optimal tuning of quantum dot energy levels. It has been also recently predicted that
efficient Cooper-pair splitting systems coupled to resonators may have a potential use as quantum heat engines. \cite{Mantovani2019Nov}
Considering all the above, there is a strong motivation to
search for novel applications of hybrid nanostructures,
involving quantum dots and superconductors,
as well as to look for different mechanisms that could optimize the Cooper pair splitting properties.

One of such mechanisms can be associated with quantum interference effects.
Indeed, both the experimental results and theoretical predictions
suggest an important role of quantum interference,
which are generally present in CPS systems.\cite{Fulop2015}
To capture and understand the role of quantum interference, a three-site model has been put forward,
where the central part is coupled to superconductor,
while the left and right arms of the splitter are modeled by remaining two separate sites.\cite{Fulop2015, Yeyati2016, bocian2018}
The transport properties of such three-site models are however still rather unexplored.
The purpose of this paper is therefore to advance further the understanding
of the Andreev transport in CPS based on triple quantum dots.
Moreover, motivated by a theoretical proposal, in which ferromagnetic contacts act as an entanglement witness
in spin transport experiments,\cite{Klobus2014} we consider a device with ferromagnetic electrodes.
Systems involving magnetic electrodes have already gained certain attention and some
spin-resolved transport aspects have been studied both theoretically and experimentally. \cite{Fabian2016Nov, Busz2017Aug, Stroganov2017Nov, Ouassou2017May}
Here, we especially focus on the analysis of the statistics of Andreev processes.
In particular, we analyze the cross-correlations between the currents flowing through
the left and right ferromagnetic junctions.
\cite{BlanterButtiker2000,Wrzesniewski2017cc,Trocha2018}
Strong positive cross-correlations are predicted in the regimes where the currents flowing
through the left and right junctions are mutually supporting each other,
which indicates high splitting efficiency of the Cooper pairs.
On the other hand, negative cross-correlations are a signature of transport processes in opposite directions,
which is an undesirable feature in the CPS devices.
Moreover, we support our analysis with the investigations of the electron waiting time distribution (WTD).
\cite{Belzig2005,Brandes2008, Ptaszynski2017Jan, Ptaszynski2017Jul}
Waiting time quantifies the time between subsequent tunneling events.
A consequent distribution can be used to characterize the CPS system,
to indicate fast and slow transport processes and give signatures of efficient Cooper pair splitting.\cite{Walldorf2018}
It is also important to note that waiting time distributions for electrons are already accessible
in experiments on quantum dot systems.\cite{Gustavsson2009, Maisi2014Jan, Gorman2017}
All the aforementioned transport characteristics are studied in two specific gate voltage detuning schemes:
in a symmetric and in an antisymmetric one.

This paper is organized as follows:
The description of theoretical framework can be found in Sec.~\ref{Theoretical description}.
Section~\ref{Results} presents the results and relevant discussion
on the cross-correlations and waiting time distribution.
Finally, the work is concluded in Sec.~\ref{Conclusions}.

\section{Theoretical description}\label{Theoretical description}

\subsection{Microscopic model}

\begin{figure}[t]
\centering
\includegraphics[width=\linewidth]{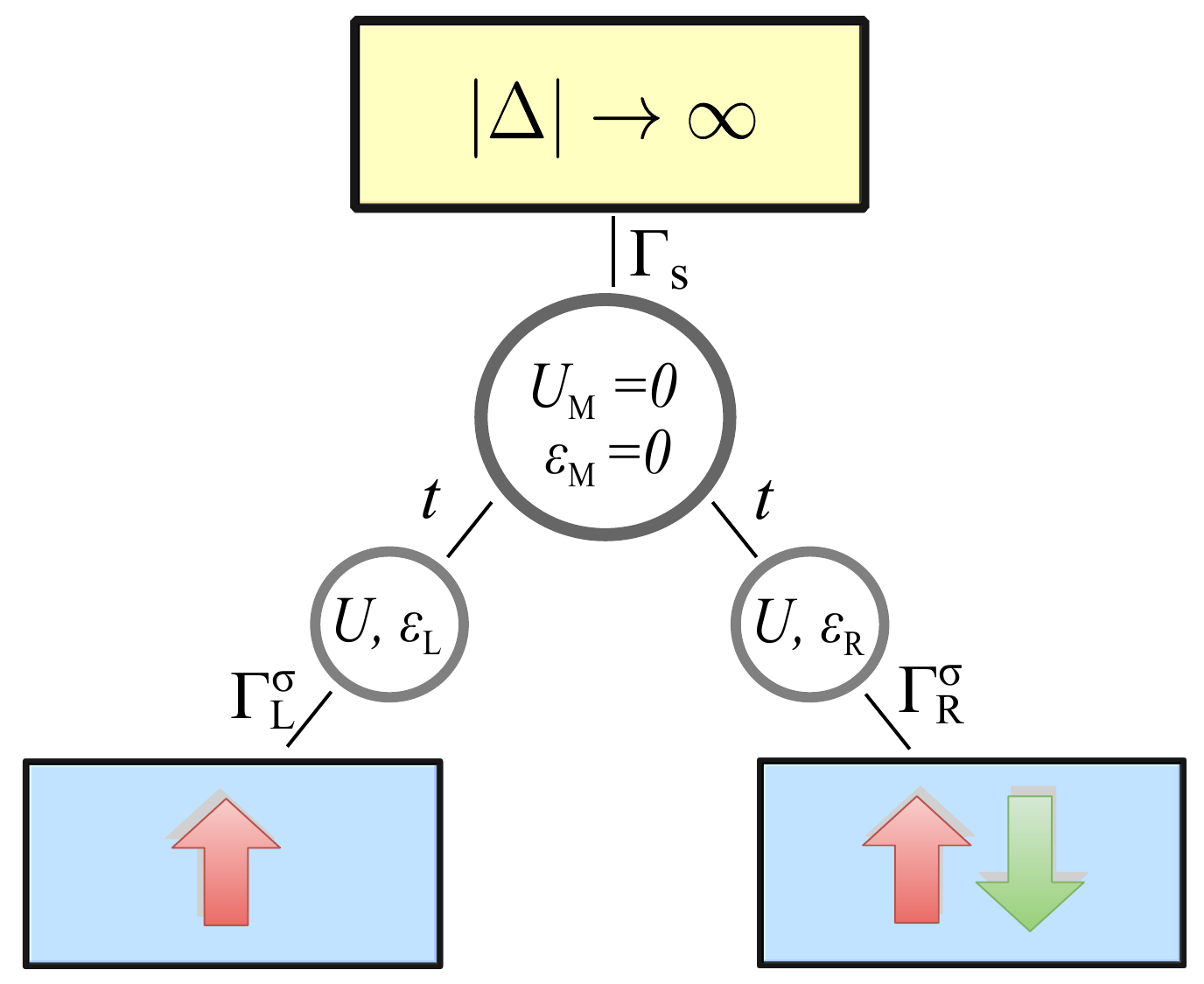}
\caption{Schematic of the considered triple quantum dot-based Cooper pair splitter.
	A large, middle quantum dot with on-site energy $\e_{\rm M}$
	is coupled directly to the superconductor
	with coupling strength $\Gamma_{\rm S}$.
	The two other quantum dots, described by on-site energies $\e_{\rm L}$ and $\e_{\rm R}$
	and Coulomb correlations $U$, respectively,
	are placed in the two arms of the splitter.
	These dots are coupled through the hopping
	matrix elements $t$ to the large dot
	and to the ferromagnetic contacts
	with the corresponding coupling strengths $\Gamma_{\rm L}^\sigma$
	and $\Gamma_{\rm R}^\sigma$. The magnetizations of ferromagnetic contacts
	are assumed to form either
	parallel or antiparallel configuration.
}
\label{fig:scheme}
\end{figure}

The considered triple quantum dot-based Cooper pair splitter
is schematically presented in Fig. \ref{fig:scheme}.
The middle quantum dot ($QD_{\rm M}$) is coupled with left ($QD_{\rm L}$)
and right ($QD_{\rm R}$) quantum dots by the hopping term $t$.
Furthermore, $QD_{\rm M}$ is attached to the s-wave superconductor (top)
with coupling strength $\Gamma_{\rm S}$, while $QD_{\rm L}$ ($QD_{\rm R}$) is coupled to left (right) ferromagnetic (FM)
electrode with spin-dependent coupling strength $\Gamma^{\sigma}_j$.
In our considerations, the magnetic moments of ferromagnets are assumed to form either parallel or antiparallel configuration.

The system is described by the total Hamiltonian:
\begin{eqnarray}\label{Eq:TotalHamiltonian}
    H = H_{\rm FM} + H_{\rm S} + H_{\rm TQD} + H_{\rm T} + H_{\rm TS}.
\end{eqnarray}
The term $H_{\rm FM}$ describes noninteracting electrons in left ($j={\rm L}$) and right ($j={\rm R}$) ferromagnetic electrodes,
\begin{eqnarray}\label{Eq:FMLeadsHamiltonian}
  H_{\rm FM} &=&
  \sum_{j={\rm L,R}}\sum_{k\sigma}\varepsilon_{jk\sigma} c^\dagger_{jk\sigma} c_{jk\sigma}
  \,,
\end{eqnarray}
with $c^\dagger_{jk\sigma}$ ($c_{jk\sigma}$) being the creation (annihilation)
operator of an electron in the $j$th lead, with momentum $k$,
spin $\sigma$ and energy $\varepsilon_{jk\sigma}$.

The second term expresses the s-wave superconductor with the mean-field BCS Hamiltonian
\begin{eqnarray}\label{Eq:SCLeadHamiltonian}
  H_{\rm S} &=&
  \sum_{k \sigma}\varepsilon_{{\rm S}k\sigma} c^\dagger_{{\rm S}k\sigma} c_{{\rm S}k\sigma} + \Delta\sum_{k}(c_{{\rm S}k\downarrow}c_{{\rm S}-k\uparrow}+ H.c.)
  \,,
\end{eqnarray}
where $c^\dagger_{{\rm S}k\sigma} (c_{{\rm S}k\sigma})$ stands for the creation (annihilation) operator of an electron in the superconductor,
with momentum $k$, spin $\sigma$ and energy $\varepsilon_{Sk\sigma}$,
while $\Delta$ is the order parameter of the superconductor, which is assumed to be real and positive.

The third term of the Hamiltonian $H$ describes the isolated triple quantum dot (TQD) and is given by
\begin{eqnarray}\label{Eq:TQDHamiltonian}
  H_{\rm TQD}&=&
  \sum\limits_{j=\rm L,M,R}\varepsilon_{j} n_{j}
  + U\sum\limits_{j={\rm L, R}} n_{j\uparrow}n_{j\downarrow}
  \nonumber\\
  &+& t\sum\limits_{j={\rm L,R}} \sum\limits_\sigma(d^\dagger_{j\sigma}d_{{\rm M}\sigma}+
  d^\dagger_{{\rm M}\sigma}d_{j\sigma})
 \,.
\end{eqnarray}
The occupation number operators are defined as follows: $n_j=n_{j\uparrow} + n_{j\downarrow}$
and $n_{j\sigma}=d^\dagger_{j\sigma}d_{j\sigma}$ with $d^\dagger_{j\sigma}(d_{j\sigma})$
being the creation (annihilation) operator of an electron in the dot $j$ with spin $\sigma$ and energy $\varepsilon_{j}$.
The last two terms of the total Hamiltonian describe the corresponding tunneling processes.
The tunneling between triple quantum dot subsystem
and ferromagnetic electrodes is modeled by
\begin{eqnarray}\label{Eq:TunnelHamiltonian}
  H_{\rm T} &=&
  \sum_{j={\rm L, R}}\sum_{k\sigma} (V^j_{k\sigma} c^\dagger_{jk\sigma}d_{j\sigma}
  + H.c.).
\end{eqnarray}
Here, $V^j_{k\sigma}$ are the tunnel matrix elements between $j$th dot and respective ferromagnetic lead.
We assume the tunnel matrix elements to be spin and momentum independent. Then,
the FM lead-dot couplings can be expressed as
$\Gamma^{\sigma}_j = 2\pi|V^{j}|^2\rho_{j\sigma}$, where $\rho_{j\sigma}$ is the
spin-dependent density of states for spin $\sigma$ of the ferromagnetic lead $j$.
The FM coupling strengths depend on the spin polarization
$p_j=(\rho_{j+} - \rho_{j-})/(\rho_{j+} + \rho_{j-})$ of the given lead
and can be denoted as $\Gamma^\pm_{j}=\Gamma_j(1\pm p_j)$ for the spin majority ($\sigma=+$) or minority ($\sigma=-$) subband.
We assume that $\Gamma_j = (\Gamma^+_j + \Gamma^-_j)/2$ and $\Gamma_{\rm L}=\Gamma_{\rm R}\equiv \Gamma$.
We also assume equal spin polarization of both ferromagnetic leads $p_{\rm L}=p_{\rm R}\equiv p$.

Finally, the tunneling between triple quantum dot and the superconductor is described by
\begin{eqnarray}\label{Eq:SCTunnelHamiltonian}
  H_{\rm TS}&=& \sum_{k\sigma} (V_{\rm S} c^\dagger_{{\rm S}k\sigma}d_{{\rm M}\sigma} + H.c.),
\end{eqnarray}
where $V_{\rm S}$ denotes the corresponding tunneling amplitude.
The superconducting lead-dot coupling is given by
$\Gamma_{\rm S} = 2\pi|V_{\rm S}|^2\rho^{\rm S}$,
where $\rho^{\rm S}$ is the density of states of superconductor in the normal state.
In further considerations the infinite superconducting energy gap limit, $\Delta \rightarrow \infty$, is assumed,
which allows us to use an effective Hamiltonian
$H_{\rm TQD}^{\rm eff}=H_{\rm QD} + H_{\rm S} + H_{\rm TS}$
including the induced superconducting pairing term in the middle dot\cite{Rozhkov2000,Domanski2016Mar}
\begin{equation}\label{Eq:Hamiltonian}
H_{\rm TQD}^{\rm eff}=H_{\rm TQD} - \frac{\Gamma_{\rm S}}{2}
(d^\dagger_{{\rm M}\uparrow} d^\dagger_{{\rm M}\downarrow} + d_{{\rm M}\downarrow} d_{{\rm M}\uparrow}).
\end{equation}

We examine the case when the bias voltage between the superconductor and FM leads
is applied in the following way:
The superconducting lead is grounded ($\mu_{\rm S}=0$),
while both left and right ferromagnetic leads amass the same electrochemical potential,
$\mu_{\rm L}=\mu_{\rm R}=eV$. Our considerations are focused on the subgap transport regime,
where the current is mediated purely by the Andreev reflection processes.
Therefore, we assume that the Coulomb interaction energy on the left and right dots
and the superconducting energy gap are the highest energies in the problem,
$\Delta, U \gg |eV|, \varepsilon_{\rm L/R}$.
On the other hand, for the middle quantum dot we assume vanishing Coulomb interaction, $U_{\rm M}=0$,
due to large size of this dot and screening effect of the superconductor.
A similar model and parameter regime was successfully applied to model recent experimental observations
on a Cooper pair splitter in external magnetic field. \cite{Fulop2015}

\subsection{Real-time diagrammatic technique}

The main aim of this paper is to study the Andreev transport properties of the considered triple quantum-dot based Cooper pair splitter.
To achieve this goal, we use the real-time diagrammatic technique, which is
based on a systematic perturbation expansion of the reduced density
matrix and the current operator with respect to the lead-dot coupling $\Gamma$.
\cite{Schoeller1994, Konig1996, Thielmann2005, Governale2008}
Here, we perform the expansion assuming a weak coupling $\Gamma$ of the triple quantum-dot with corresponding ferromagnetic lead,
while the coupling to the superconductor $\Gamma_{\rm S}$ is treated exactly.
Consequently, in our calculations, we take into account the lowest-order tunneling processes,
namely sequential tunneling through the ferromagnetic junctions.

The self-energies corresponding to the quantities of interest
are calculated by summing up the contributions of the relevant
diagrams determined with the use of respective diagrammatic rules.
\cite{Schoeller1994, Konig1996, Thielmann2005, Governale2008}
Furthermore, solving the kinetic equation
\begin{eqnarray} \label{Eq:KineticEq}
  \textbf{Wp}^{\rm st}&=& 0
\end{eqnarray}
allows one to obtain the vector of occupation probabilities ${p}^{\rm st}_\chi$
of the eigenstates of the effective Hamiltonian $H_{\rm TQD}^{\rm eff}$,
$H_{\rm TQD}^{\rm eff}|\chi \rangle=\varepsilon_{\chi}|\chi \rangle$.
The matrix elements $W_{\chi' \chi}$ of \textbf{W} are
the self-energies corresponding to the transitions between the states $|\chi' \rangle$ and $|\chi \rangle$.
Details concerning the calculation of transition rates $W_{\chi' \chi}$ are described in the Appendix.

The current flowing through the ferromagnetic junction ${\it j}$ can be found from
\cite{Thielmann2005}
\begin{eqnarray} \label{Eq:Current}
  I_j &=& \frac{e}{2\hbar} {\rm Tr}\left \{ \textbf{W}^{I_j}{\mathbf p}^{\rm st} \right \}
 \,,
\end{eqnarray}
where the matrix $\textbf{W}^{I_j}$ takes into account the number of transferred electrons through the junction $j$.
The current flowing through the superconducting junction
$I_{\rm S}$ can be found by means of the Kirchhoff's law, $I_{\rm S}=I_{\rm L} + I_{\rm R}$,
while the corresponding differential conductance is evaluated from,
$G_{\rm S} = dI_{\rm S}/dV$.

Having the Andreev current calculated in the case of parallel and antiparallel magnetic configuration of the ferromagnetic leads,
the tunnel magnetoresistance (TMR) can be found from \cite{Futterer2009,Weymann2014}
\begin{eqnarray} \label{Eq:TMR}
  {\rm TMR} &=& \frac{I_{\rm S}^{AP}-I_{\rm S}^{P}}{I_{\rm S}^{P}}
 \,.
\end{eqnarray}
The foregoing definition of the TMR is different to the one known from the Julliere model \cite{Julliere1975}
and the modification is justified by the fact that in hybrid systems with superconducting and two ferromagnetic electrodes,
the Andreev current in the antiparallel configuration
is usually greater than that in the parallel alignment. \cite{Futterer2009,Weymann2014, Trocha2015}

The calculations of the zero-frequency current-current cross-correlations require the evaluation
of the self-energy matrices for both the left and right junctions,
i.e. ${\bf W}^{I_{\rm L}}$ and ${\bf W}^{I_{\rm R}}$.
The correlation function in the sequential tunneling approximation
has contributions only from the current operators appearing in two distinct irreducible blocks.
\cite{Wrzesniewski2017cc,Trocha2018}
Therefore, the following expression allows for calculating the
cross-correlation function between currents flowing through the left and right junctions
\cite{Thielmann2005}
\begin{eqnarray} \label{Eq:Slr}
  S_{\rm LR}=\frac{e^2}{\hbar} {\rm Tr} \left\{({\mathbf W}^{I_{\rm L}}\mathbf{P} {\mathbf W}^{I_{\rm R}}+{\mathbf W}^{I_{\rm R}}\mathbf{P} {\mathbf W}^{I_{\rm L}}){\mathbf p}^{st}\right\}
 \,.
\end{eqnarray}
The propagator ${\mathbf P}$ can be found from ${\bf \tilde{W}P=p^{\rm st}e^{T}-1}$,
where the matrix ${\bf \tilde{W}}$ is similar to ${\bf W}$,
but with one row substituted by $(\Gamma, \Gamma, ... , \Gamma)$, and ${\bf e^T}=(1,1, ..., 1)$.

\subsection{Waiting time distribution}

We enrich the analysis of the transport properties of the considered system
with the study of electron waiting time distributions. \cite{Brandes2008,Walldorf2018}
The electron waiting time is a time elapsed between two subsequent physical
events of a fixed type. In our calculations, we consider tunneling jumps
of the electron with spin $\sigma$ through the $j={\rm L,R}$ junction.
Having set a sequence of two consecutive tunneling events that we are interested in evaluation,
the waiting time can be understood in a following way.
When the tunneling of the first type takes place, the measurement of elapsed time is started.
Then the system evolves until the time measurement is stopped by the occurrence
of the subsequent tunneling event. Experimentally, the tunneling of electrons
can be measured in real-time with charge detectors. \cite{Maisi2014Jan, Gorman2017}

In order to calculate the relevant waiting times,
we assume negatively biased ferromagnetic leads in the way
that relevant quantum dots' levels are deep in the transport window.
Then, tunneling has a unidirectional character and the system can be described
with the aid of Markovian quantum master equation for the reduced density matrix $\hat{\rho}$
\cite{Kossakowski1972Dec,Gorini1976May,Lindblad1976Jun,Breuer2007Jan}
\begin{equation} \label{Eq:Lindblad}
\frac{d}{dt}\hat{\rho}=\mathcal{L}\hat{\rho}=-i[H_{\rm TQD}^{\rm eff}, \hat{\rho}] + \mathcal{D}\hat{\rho}.
\end{equation}
The Liouvillian $\mathcal{L}$ consists of coherent processes described by $H_{\rm TQD}^{\rm eff}$ and
incoherent electron jumps described by the Lindblad dissipator \cite{Gorini1976May,Lindblad1976Jun}
\begin{equation} \label{Eq:Dissipator}
    \mathcal{D}\hat{\rho}=\sum_{j=\rm L,R}\sum_\sigma \Gamma_j \left [ d_{j \sigma} \hat{\rho} d^\dagger_{j \sigma}-
    \frac{1}{2} \left \{d^\dagger_{j \sigma}d_{j \sigma},\hat{\rho}\right \} \right ].
\end{equation}
Note that the density matrix contains both diagonal and off-diagonal matrix elements
related to coherences between states.

The waiting time distribution between transitions of type $a$ and $b$ can be then expressed as
\cite{Brandes2008}
\begin{eqnarray} \label{Eq:wtd}
    \mathcal{W}_{ab}(\tau)=\frac{{\rm Tr}[\mathcal{J}_{a} e^{(\mathcal{L}-\mathcal{J}_{a})\tau} \mathcal{J}_{b} \hat{\rho}_{st}]}{{\rm Tr}[\mathcal{J}_{b} \hat{\rho}_{st}]},
\end{eqnarray}
where $\hat{\rho}_{st}$ is the stationary density matrix and the jump superoperators are
found from $\mathcal{J}_{a} \hat{\rho}=\Gamma_a d_{a} \hat{\rho} d^\dagger_{a}$,
where $a=j \sigma$ and $b=j' \sigma'$ denote two different tunneling acts through the
$j$ and $j'$ junctions of electrons with spins $\sigma$ and $\sigma'$.

\section{Results and discussion}\label{Results}

We focus on the analysis of the subgap transport where the CAR processes are expected to be dominant. Therefore, we assume the limit of infinite intradot Coulomb correlations on the left ($QD_{\rm L}$) and right ($QD_{\rm R}$) dots, i.e. $U \rightarrow \infty$. As a result, double occupation on each of the side dots is forbidden. The Cooper pairs are injected from the superconductor onto the central dot with vanishing intradot interaction $U_{\rm M}=0$ and orbital energy level $\varepsilon_{\rm M}=0$. Subsequently, the strong intradot Coulomb correlations present in the side dots are enforcing electron pair splitting and transport through CAR processes. The other parameters are set as follows: $\Gamma_{\rm S}\equiv1$ is used as energy unit,
interdot hopping is $t=1$ and the ferromagnetic lead spin polarization is $p=0.5$.
The temperature is equal to $T=0.02$. The electrons in the ferromagnetic leads
are described by the Fermi-Dirac distribution function
$f_{\rm L/R}(\omega)=1/(e^{(\omega-\mu_{\rm L/R})/T}+1)$, with $k_B\equiv 1$ and $\mu_{\rm L/R}$ denoting
the chemical potential of the left/right lead.

In order to thoroughly analyze the transport properties of the considered system, we examine two distinct gate detuning protocols.
In the symmetric case, the gate voltages associated with the left
and right quantum dots are varied in a symmetric way, i.e. $\varepsilon=\varepsilon_{\rm L}=\varepsilon_{\rm R}$.
On the other hand, for the antisymmetric case, we detune the levels as follows
$\varepsilon=\varepsilon_{\rm L}=-\varepsilon_{\rm R}$.
Finally, we recall that the superconducting electrode is always grounded,
while equal potential $\mu_{\rm L}=\mu_{\rm R}=eV$
is applied to both ferromagnetic leads.

\subsection{Andreev current, differential conductance and TMR}
\subsubsection{Symmetric gate detuning}

\begin{figure}[t]
\centering
\includegraphics[width=\linewidth]{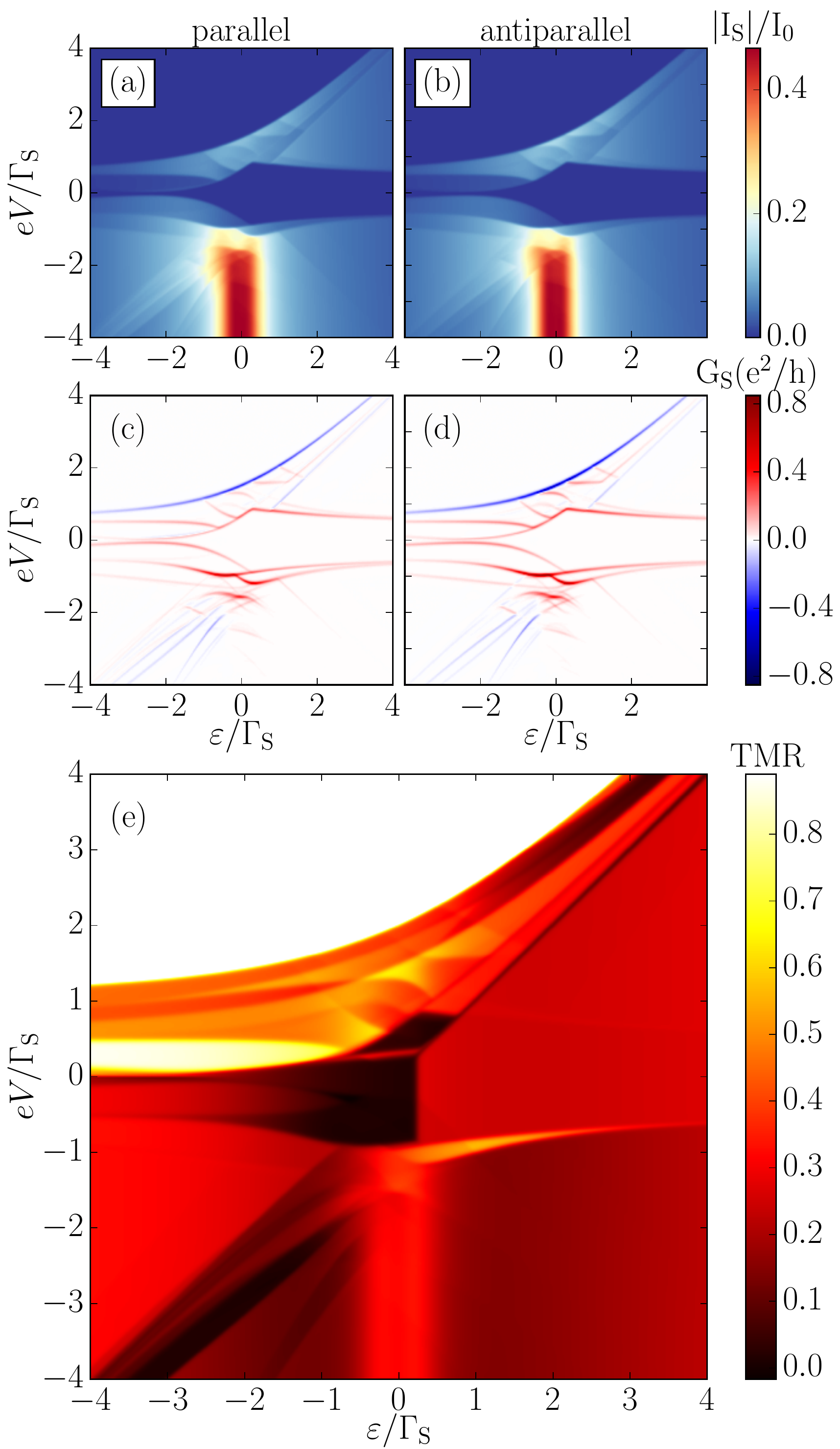}
\caption{The absolute value of the Andreev current $I_{\rm S}$ [(a) and (b)]
	and the Andreev differential conductance $G_{\rm S}$ [(c) and (d)] as a function of
	symmetric detuning $\varepsilon\equiv \varepsilon_{\rm L}=\varepsilon_{\rm R}$
	and the applied bias voltage $eV$.
	The left column shows the results for the parallel magnetic configuration,
	while the right column corresponds to the antiparallel one.
	The tunnel magnetoresistance (TMR) dependence on the detuning and applied bias voltage is shown in panel (e).
	The parameters are: $\Gamma_S\equiv1$, used as the energy unit,
	$\Gamma=0.1$, $t=1$, $T=0.02$, $p=0.5$
	and $I_0=e\Gamma/\hbar$.}
\label{fig:current_symm}
\end{figure}

Figure \ref{fig:current_symm} displays the dependence of the Andreev current,
differential conductance and the TMR on the applied bias voltage $eV$
and the symmetric dot-level detuning $\varepsilon\equiv \varepsilon_{\rm L}=\varepsilon_{\rm R}$.
The left column [panels (a) and (c)] concerns the parallel magnetic configuration,
while the right column [panels (b) and (d)] shows the results in the case of the antiparallel configuration.
The resulting TMR is shown in Fig. \ref{fig:current_symm}(e).

Let us first discuss the Andreev current characteristics.
Generally, in the low bias voltage regime the current is strongly suppressed.
The triple quantum dot is in the singlet ground state,
in which the wave function is distributed over all three dots.
When the system is detuned towards negative energies,
$\varepsilon<0$,
the blockade regime becomes relatively narrow ($eV/\Gamma_{\rm S}<0.1$).
Such detuning enhances the electron occupation of the side dots
$QD_{\rm L}$ and $QD_{\rm R}$ at the cost of the $QD_{\rm M}$'s occupation.
This configuration results in an excited state,
which is energetically close to the ground state, and when the small bias voltage is applied,
the current start to flow by continuous transitions
between the two-electron singlet ground state and one-electron excited doublet.
The low energy position of the side dots also influences
the intensity of the Andreev current,
which is higher when the applied bias is negative and particles
flow from the superconductor into ferromagnetic leads
than in the case of the opposite voltage and current direction.
Interestingly, for the opposite values of detuning,
when $\varepsilon>0$,
the blockade is present for wider range of bias, $-1/\sqrt{2} < eV/\Gamma_{\rm S} <1/\sqrt{2}$,
as compared to the previously discussed case due to the increased
stability of singlet state on the central dot.
The considered low-bias behavior is also reflected in the Andreev differential conductance.
The regime of the current blockade is clearly placed between the two
nearly-parallel lines of maxima associated with relevant states
entering the transport window. The maximum value is higher in the antiparallel configuration
than in the parallel one, which also results in higher current $I_{\rm S}^{AP}>I_{\rm S}^{P}$,
and eventually positive TMR.

Furthermore, if we examine a higher bias voltage regime, $|eV/\Gamma_{\rm S}|>1$,
there are two pronounced features visible in transport characteristics.
The first one, when the system is negatively biased ($eV/\Gamma_{\rm S}<-1$)
and Cooper pairs are extracted from the superconductor,
is the regime of maximal Andreev current.
The side dots need to be tuned near the Fermi level of the superconductor,
$\varepsilon=0$. Inconveniently, the plateau is in relatively narrow range of detuning parameter,
with half of its maximum value when detuned to approximately $|\varepsilon|/\Gamma_{\rm S}\approx 0.5$.

When the system is biased in the opposite direction,
a strong negative differential conductance (NDC) is present,
starting from relatively small voltages ($eV/\Gamma_{\rm S} \approx 0.7$) for deep energy levels of the side dots
($\varepsilon/\Gamma_{\rm S}<-2$) and spanning to higher bias while detuning toward positive values.
Further increase of the bias voltage to overcome the NDC
leads to the regime of Andreev current suppression
due to a triplet blockade present in both magnetic alignments.
This effect is well known from the analysis of transport through the double quantum-dot splitters.
\cite{Eldridge2010, Trocha2015}
Here, we want to note that if the double occupation of side dots is allowed by setting
finite intradot Coulomb interactions, the triplet blockade can be lifted by applying
high enough bias voltage. This regime is however difficult to explore experimentally,
as most often the charging energy is higher than the energy gap $\Delta$
of many currently known superconductors.\cite{DeFranceschi2010,Lee2014}
Moreover, the differential conductance plots are exposing
numerous additional local extrema, which are resulting from the complex
electronic structure of the triple quantum-dot system,
however, they do not lead to qualitatively new effects.

Due to the general property of the system that the
Andreev current is higher in antiparallel configuration than in the parallel one,
i.e. $I_{\rm S}^{AP}>I_{\rm S}^{P}$, the resulting TMR is positive
in wide range of transport parameters.
The extensive region for positive bias marked by white color is associated with the presence of the triplet blockade,
where Andreev currents in both magnetic configurations are strongly suppressed. In consequence, the TMR becomes indeterminate.
Other than that, the tunnel magnetoresistance has a positive and moderate values for wide range of parameters,
with a considerable raise for the earlier discussed regime of maximal Andreev current
($\varepsilon=0$ and $eV/\Gamma_{\rm S}<0$), where ${\rm TMR}=p^2/(1-p^2) =1/3$ (for assumed $p=0.5$)
and for low positive bias ($0<eV/\Gamma_{\rm S}<0.5$) and negative detuning,
where TMR becomes enhanced and reaches ${\rm TMR \approx 0.88}$.

\subsubsection{Antisymmetric gate detuning}

\begin{figure}[t]
\centering
\includegraphics[width=\linewidth]{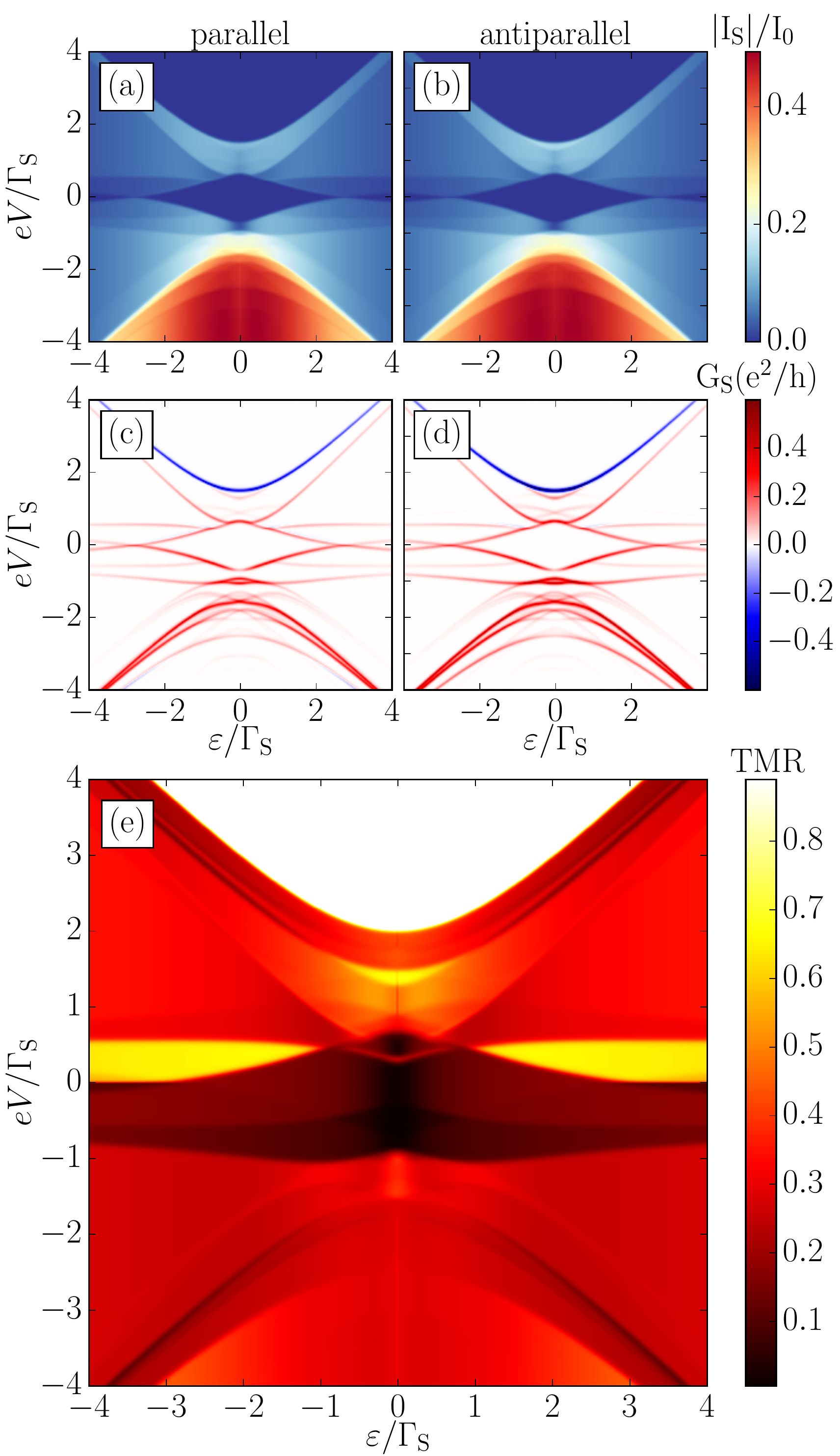}
\caption{The absolute value of the Andreev current [(a) and (b)] and the
	Andreev differential conductance [(c) and (d)] as a function of antisymmetric
	detuning $\varepsilon\equiv \varepsilon_{\rm L}=-\varepsilon_{\rm R}$
	and applied bias voltage $eV$. The left column shows the results for the parallel magnetic configuration,
	while right column is for the antiparallel configuration.
	The TMR dependence on detuning and applied bias voltage is shown in (e).
	The parameters are the same as in Fig. \ref{fig:current_symm}.}
\label{fig:current_asymm}
\end{figure}

Let us now consider a different scheme of detuning the energy levels of the system---the antisymmetric one, where
$\varepsilon\equiv \varepsilon_{\rm L}=-\varepsilon_{\rm R}$.
Such detuning results in symmetric transport dependencies
with respect to the change of sign of detuning $f(\varepsilon)=f(-\varepsilon)$,
where $f=I_{\rm S}$, $G_{\rm S}$, ${\rm TMR}$ and $S_{\rm LR}$.
This detuning protocol is expected to be favorable for high
CPS conductance. \cite{DLoss2001, Yeyati2016}
The energy conservation holds when the Cooper pair
is transferred from the middle dot $QD_{\rm M}$ with energy $\varepsilon_{\rm M}=0$,
through the side dots, where due to detuning scheme $\varepsilon_{\rm L}=-\varepsilon_{\rm R}$, the total energy can be approximated by $\varepsilon_{\rm L} + \varepsilon_{\rm R} = 0$,
which implies high Andreev current.
However, it is important to note that the
eigenenergies are also influenced by other parameters of the system, for instance by the hopping $t$,
therefore the above discussion provides rather qualitative picture,
but the quantitative numerical results support it.

Figure \ref{fig:current_asymm} displays the Andreev current,
differential conductance and TMR as a function of the applied bias $eV$
and the antisymmetric detuning $\varepsilon$.
The low bias voltage regime again exposes
the current blockade for $|\varepsilon/\Gamma_{\rm S}|<2$ and $|eV/\Gamma_{\rm S}|<1$.
With further increase of the bias voltage, excited stats enter the transport window
and the Andreev current starts to flow.
For the positive bias, $eV/\Gamma_{\rm S}>2$, the triplet blockade is present as well,
however, the shape of the region with suppressed Andreev current
is different than in the case of symmetric detuning
and the associated minimum in Andreev differential conductance is now sharper.
For the system strongly biased in the opposite direction,
with electrons flowing into the ferromagnetic leads,
there is a significantly wider range of the regime with high Andreev current,
where $|I_{\rm S}|/I_0\gtrsim 0.4$.
This characteristic clearly shows that
antisymmetric detuning is advantageous for the flow of high Andreev current.

On the other hand, the behavior of the TMR is rather similar
to the symmetric detuning case. For the whole parameter space the TMR is positive.
The undetermined range for positive bias due to the triplet blockade (white region) is present likewise.
Moreover, there is a low-bias regime for a relatively wide range of detuning
($|\varepsilon/\Gamma_{\rm S}|>1.5$) where the ${\rm TMR}=2p^2/(1-p^2)=2/3$.

It is important to note that in the antisymmetric case all the interesting regimes
with different transport properties are more distinguishable
and easier to adjust than in the symmetric case.

Finally, we would like to remark that a finite Coulomb interaction on the central dot $U_{\rm M}>0$
has only a moderate influence on the presented results as long as it is smaller than the applied bias voltage.
For both considered gate detuning protocols the maximal values of the Andreev current become slightly reduced for finite $U_{\rm M}$.
Moreover, in the case of symmetric detuning, the regime of maximal Andreev current is shifted towards negative detuning,
while for antisymmetric one, the low-bias blockade region becomes enlarged.

\subsection{Current cross-correlations}

\begin{figure}[t]
\centering
\includegraphics[width=\linewidth]{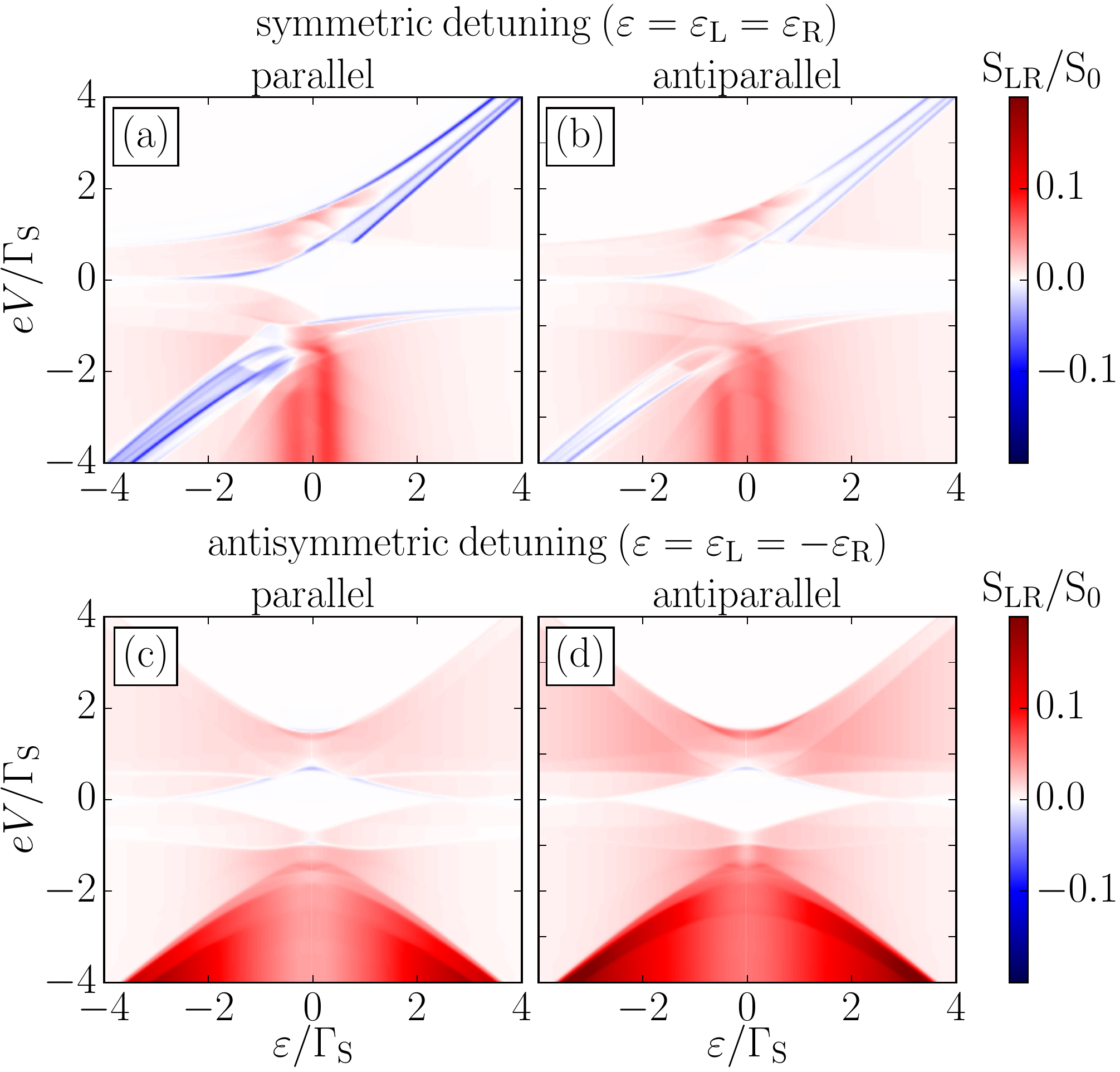}
\caption{The current-current cross-correlations as a function of detuning $\varepsilon$ and the applied bias voltage $eV$.
The top row corresponds to the case of symmetric detuning,
$\varepsilon\equiv \varepsilon_{\rm L}=\varepsilon_{\rm R}$, while the bottom row is calculated for antisymmetric detuning,
$\varepsilon\equiv \varepsilon_{\rm L}=-\varepsilon_{\rm R}$.
The left column shows the results for the parallel magnetic configuration and the right column is for the antiparallel alignment.
The parameters are the same as in Fig. \ref{fig:current_symm} and $S_0=e^2\Gamma/\hbar$.}
\label{fig:slr}
\end{figure}

The cross-correlations in Cooper pair splitters between currents flowing through
the left and right junctions indicate important transport features.
\cite{Wrzesniewski2017cc, Trocha2018}
Primarily, the strong positive cross-correlations are associated with mutually
supporting transport processes between the left and right junctions.
In the case of Cooper pair splitters,
this feature is also a signature of high splitting efficiency.
On the other hand, the presence of negative cross-correlations is associated with
the transport processes taking place in opposite directions.
The current cross-correlations for the considered model,
in the case of both symmetric and antisymmetric detuning are shown in Fig. \ref{fig:slr}.

The general result is that, analogically to a single and double quantum-dot based Cooper pair splitters with ferromagnetic leads \cite{Wrzesniewski2017cc, Trocha2018}, in the antiparallel magnetic configuration
the positive cross-correlations have higher values,
while the negative ones are diminished, which is contrary to the case of the parallel configuration.
This is associated with the strong influence of the spin-dependent tunneling on the Andreev transport.
This effect can be tuned by changing the value of leads spin polarization $p$,
where along with increasing $p$, positive cross-correlations become enhanced
in the antiparallel configuration, while the negative values are increased in the parallel one.

In the case of symmetric detuning of the system,
see Figs.~\ref{fig:slr}(a) and (b),
the main area of strong positive cross-correlations coincides
with the maximal Andreev current.
However, in the vicinity of $\varepsilon=0$ for applied negative bias and when the quantum-dot system is detuned further above $|\varepsilon/\Gamma_{\rm S}|>1$,
the cross-correlations become significantly diminished.
Another striking feature in the considered case is the presence of strong anti-diagonal minima spanning in the discussed plots,
evolving around the resonances between detuning and chemical potential of the ferromagnetic leads, i.e. $eV \approx \varepsilon$.
This is the condition strongly enhancing possibility of single-electron tunneling back
and forth through both ferromagnetic junctions simultaneously,
resulting in negative values of current-current cross-correlations.
The important fact is that in the case of double quantum-dot systems,
the negative cross-correlations are present in the case of finite hopping between the dots.
Here, the left and right dots are not coupled by direct hopping, however,
the presence of the middle dot mediates this mechanism.
To sum it up, the analysis of cross-correlations for the symmetric detuning
concludes with rather unfavorable transport properties,
with weakly correlated transport in a wide range of parameters.
Large values of cross-correlations are present
only  for small range of $\varepsilon$ and negative bias, see Figs.~\ref{fig:slr}(a) and (b).
Finally, the presence of the negative cross-correlations is also an unwanted feature for optimal operation of the CPS device.

The current-current cross correlations in the case of antisymmetric detuning,
see Figs.~\ref{fig:slr}(c) and (d),
expose more attractive features as far as the Andreev transport and Cooper pair splitting are concerned.
Negative cross-correlations are almost completely washed out in both magnetic configurations.
There is a small minimum formed at the edge of the low-bias
current blockade for the positive bias voltage.
However, for the current flowing in the opposite direction,
i.e. when the Cooper pairs are extracted from the superconductor
and electrons are transported into ferromagnetic leads,
there are no negative cross-correlations in the whole range of detuning parameter $\varepsilon$
for $eV<0$. In this case, due to the nature of antisymmetric detuning,
the possibility of resonant transport in both directions
between distinct ferromagnetic leads and coupled side dots is blocked.
Moreover, the positive cross-correlations in this scheme are strongly enhanced,
as compared to the symmetric detuning protocol,
for a significant range of negative bias $eV/\Gamma_S<1$
and wide range of detuning parameter $\varepsilon$.
When transport takes place in this regime, not only the Andreev current
is high and stimulated by the energy conservation of the transferred particles,
but also high positive cross-correlations indicate efficient splitting
and mutual support of tunneling processes through left and right junctions.
Both important properties of the antisymmetric detuning scheme
are encouraging for using it in Cooper pair splitting devices,
allowing for considerable and unidirectional Andreev current
as well as optimal splitting efficiency of the emitted Cooper pairs.

\subsection{Influence of inter-dot hopping amplitude}

\begin{figure}[t]
\centering
\includegraphics[width=\linewidth]{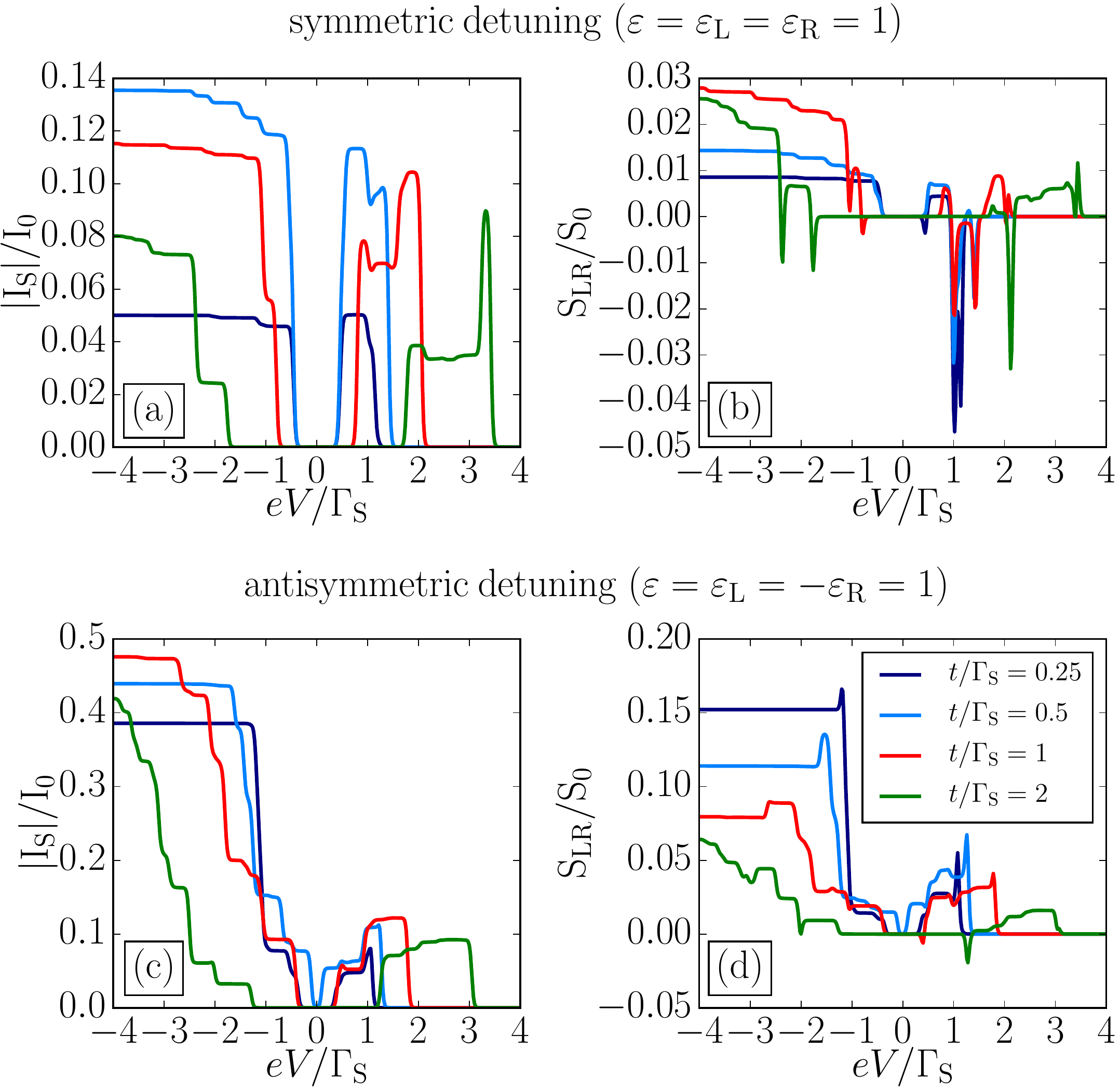}
\caption{The absolute value of the Andreev current [(a) and (c)] and current-current cross-correlations [(b) and (d)]
as a function of the applied bias voltage $eV$
for different values of inter-dot hopping $t$ in the antiparallel magnetic configuration.
The top row corresponds to the case of symmetric detuning,
$\varepsilon\equiv \varepsilon_{\rm L}=\varepsilon_{\rm R}=1$, while the bottom row is calculated for antisymmetric detuning,
$\varepsilon\equiv \varepsilon_{\rm L}=-\varepsilon_{\rm R}=1$.
The parameters are the same as in Fig. \ref{fig:current_symm}.}
\label{fig:slr_vs_t}
\end{figure}

An important factor influencing the transport properties of the multi-quantum-dot based Cooper pair splitters
is the amplitude of the inter-dot hopping $t$. As shown recently for double quantum-dot systems,\cite{Wrzesniewski2017cc}
finite hopping is responsible for the regimes with negative cross-correlations as well as attenuation of strong positive correlations.
In order to get a better understanding of the role of hopping in transport through triple quantum-dot Cooper pair splitters,
in Fig.~\ref{fig:slr_vs_t} we show the Andreev current and the corresponding cross-correlations
for the system detuned symmetrically ($\varepsilon\equiv \varepsilon_{\rm L}=\varepsilon_{\rm R}=1$)
and antisymmetrically ($\varepsilon\equiv \varepsilon_{\rm L}=-\varepsilon_{\rm R}=1$) for various hopping amplitudes $t$.
The presented results are shown for the antiparallel magnetic configuration,
in which the calculated quantities have higher values with respect to the case of parallel configuration.
For positively biased system ($eV>0$), the main effect of varying the hopping amplitude
is a change of the position of triplet blockade,
i.e. with increase of $t$ the triplet blockade emerges at higher voltages.
This effect is similar in both considered detuning schemes, see the left column of Fig.~\ref{fig:slr_vs_t}.
The dependence of current cross-correlations is congruent with this observation
as the fluctuations are enhanced for the bias voltage range preceding the current suppression due to triplet blockade.

The behavior is more interesting for the negative bias voltage ($eV<0$),
where the current flows from superconductor toward ferromagnetic leads.
For the symmetric detuning, small hopping amplitudes ($t/\Gamma_{\rm S}\leq 0.25$)
result in weak Andreev current and weak cross-correlations, see Figs.~\ref{fig:slr_vs_t}(a) and (b).
When the hopping amplitude $t$ is increased,
both considered transport quantities acquire higher values,
however for $t/\Gamma_{\rm S}> 1$ a significantly higher bias voltage
has to be applied in order to achieve maximal Andreev current.
This effect is also visible for negative bias in the antisymmetric detuning case.
As the hopping $t$ becomes stronger, the absolute value of the applied bias
needs to be higher for the Andreev current to flow through the device.
Surprisingly, the current cross-correlations expose opposite dependence
on $t$ in the antisymmetric detuning case as compared to the symmetric case.
For low values of inter-dot hopping amplitude, the positive cross-correlations are strongly enhanced,
while the Andreev current remains significant, see Figs.~\ref{fig:slr_vs_t}(c) and (d).
The form of the molecular states taking dominant part in transport
can shed some light on the observed behavior.
In general, the asymmetric detuning allows to form states
with high amplitudes of electron wavefunction on the left and right quantum dots.
Such distribution is favorable both for efficient transport and Andreev current,
as these dots are directly coupled to ferromagnetic leads,
as well as for Cooper pair splitting, which is exposed in strong positive cross-correlations.
Now, if the hopping $t$ is acquiring higher and higher values,
the distribution of the electron density for the TQD eigenstates in the transport window
is shifted toward middle quantum dot. Moreover, hopping $t$ is responsible for Andreev bound states splitting.
When an Andreev bound state is near the edge of the bias window,
its contribution to transport can be suppressed,
when the amplitude $t$ becomes significant enough to split the state out of the bias window.
This effect results in additional Coulomb steps appearing in current-voltage curves,
whilst the inter-dot hopping amplitude is increased.
As a consequence of these two effects, the Andreev current is lowered and less cross-correlated.

The above analysis demonstrates that it is favorable to tune the device asymmetrically,
with possibly small inter-dot hopping $t$,
especially when the efficient splitting is expected within low-bias voltage regime.
It is important to note, however, that the above considerations
are valid for $t\gg\Gamma$, where the molecular states on TQD system are well-developed.
Further enhancement of transport properties should not be expected for even smaller hopping amplitudes $t<\Gamma$.

\subsection{Electron waiting time distribution}

In this section, we analyze the distribution of electron waiting times
for the considered three-site Cooper pair splitter. We designate waiting times
between two specific jumps into ferromagnetic leads,
focusing especially on those related to splitting of the Cooper pairs.
\cite{Walldorf2018}
The relevant energies present in the system fulfill the following relation:
$\Delta,U \gg |eV| \gg \varepsilon_{\rm L}, \varepsilon_{\rm M}, \varepsilon_{\rm R}, \Gamma_{\rm S}, \Gamma$.
We also assume that the temperature $T$ is low, such that
transport is unidirectional and the effects of thermally activated processes and thermal smearing are negligible.
Then, the considered parameter regime allows us
to describe the transport with Markovian master equation.\cite{Gurvitz1996Jun, Gurvitz1998Mar}

The results are shown for the symmetric setup where all three
dots have the energy levels in resonance, $\varepsilon_{\rm M}=\varepsilon_{\rm L}=\varepsilon_{\rm R}=0$,
and for a strong antisymmetric detuning,
where the middle dot remains intact $\varepsilon_{\rm M}=0$,
while the side dots' energy levels are set as: $\varepsilon_{\rm L}=-\varepsilon_{\rm R}=10\Gamma_{\rm S}$.
For both detuning cases, we examine two distinct coupling regimes:
in the first one we assume coupling ratio $\Gamma=\Gamma_{\rm S}/10$,
where Cooper pairs are rapidly injected into the triple dot,
but the rate of transferring the electrons into the drains is slow.
As a result, there are coherent oscillations present in the distribution
that are taking place between the TQD and superconductor,
as well as oscillations due to the internal dynamics of the quantum dot subsystem.
In the second case, we set $\Gamma=10 \Gamma_{\rm S}$,
where the fast transfer of the emitted electrons is observed and the splitting is efficient,
while the oscillations are strongly suppressed due to fast electron transfer rate
to the ferromagnetic contacts.
We note that similar coupling considerations have been already successfully studied
in a hybrid single quantum dot system, where the shot-noise revealed regimes of strong superconducting
correlations and suppression of the proximity effect. \cite{Braggio2011Jan}

\subsubsection{Absence of gate detuning ($\varepsilon_{\rm L}=\varepsilon_{\rm R}=0$)}

\begin{figure}[t]
	\centering
	\includegraphics[width=\linewidth]{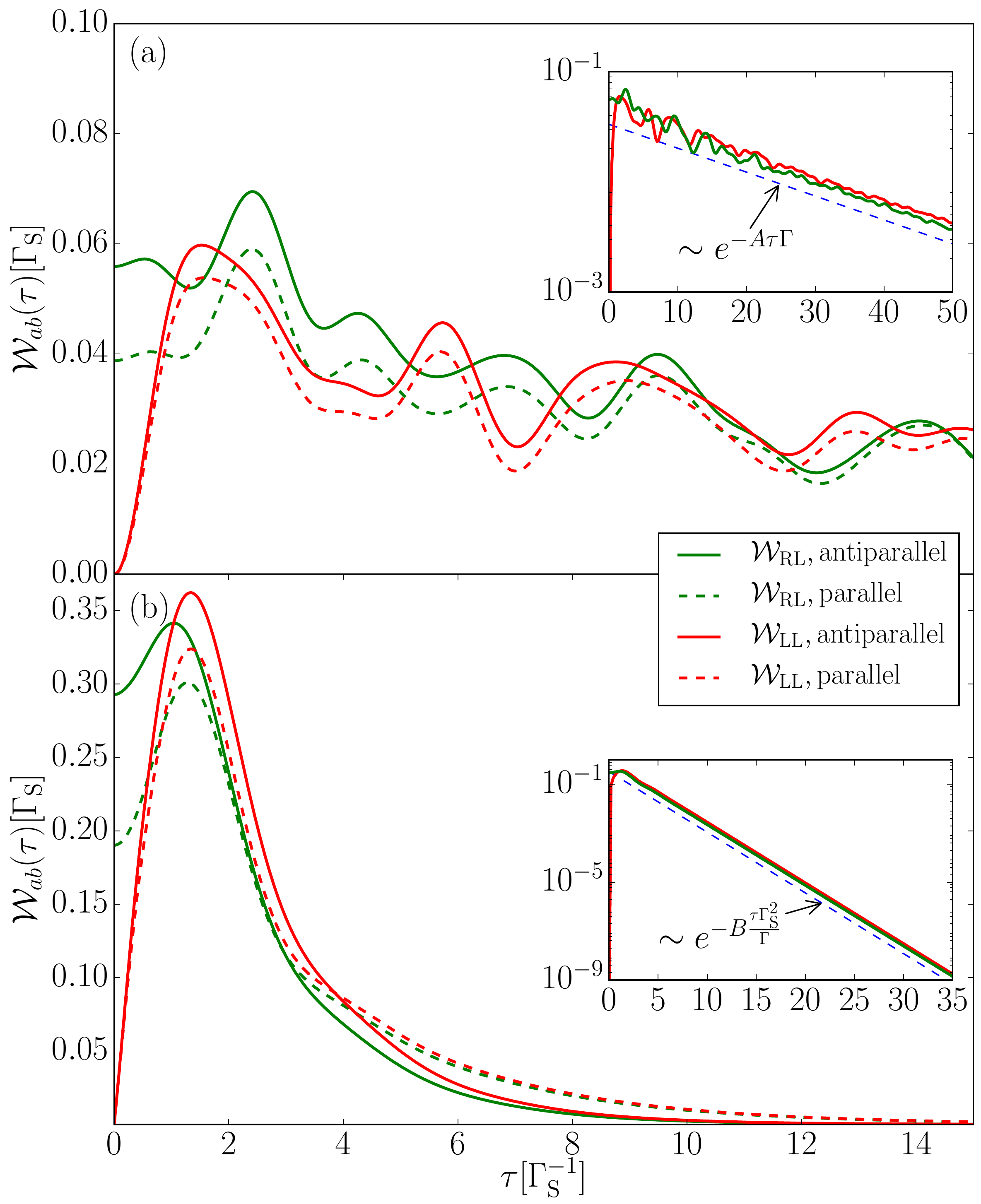}
	\caption{The charge-resolved electron waiting time distribution
		of the triple quantum-dot-based Cooper pair splitter.
		The parameters are: $\Gamma_{\rm S}\equiv 1$, $t=1$ and $\varepsilon_{\rm M}=\varepsilon_{\rm L}=\varepsilon_{\rm R}=0$.
		Panel (a) shows the results when the coupling to ferromagnetic leads is equal to
		$\Gamma=\Gamma_{\rm S}/10$, while panel (b) presents the results
		for $\Gamma=10\Gamma_{\rm S}$.
		The dashed (solid) lines correspond to the parallel (antiparallel) magnetic configuration of the ferromagnetic leads.
		The corresponding insets present the WTD plotted on logarithmic scale in an extended time period with approximated decay rates.}
	\label{fig:wtd1}
\end{figure}

Let us start the discussion with the symmetric case,
in which we assume that all three quantum dots are in resonance with the Fermi level,
i.e. $\varepsilon_{\rm M}=\varepsilon_{\rm L}=\varepsilon_{\rm R}=0$.
In Fig. \ref{fig:wtd1} we show the charge-resolved waiting time distributions for two transition types,
$\mathcal{W}_{\rm LL}$ where the WTD is estimated between two subsequent
tunneling events through the left junction, and $\mathcal{W}_{\rm RL}$,
where WTD is shown for the tunneling through the left junction
followed by the tunneling through the right one.
The WTD is considered in both the parallel (dashed lines) and antiparallel (solid lines)
magnetic configurations of the ferromagnetic leads.

Figure \ref{fig:wtd1}(a) presents the results in the case
when the coupling to ferromagnetic leads is weaker than
the coupling to superconductor, $\Gamma=\Gamma_{\rm S}/10$.
In this setup, the Cooper pairs are injected in a fast manner from superconductor to the triple dot,
while the bottleneck of the transport through the system is due to the
relatively weak coupling of TQD to the FM contacts.
In consequence, the waiting time distribution as a function of time $\tau$
exhibits an oscillatory behavior, indicating the presence of coherent oscillations of
Cooper pairs between superconductor and TQD and
those resulting from the internal dynamics of the triple quantum dot subsystem.
These oscillations are however not as regular as predicted
for single or double quantum dot systems,\cite{Rajabi2013, Walldorf2018}
which is due to the complex electronic structure and the interplay of correlations
in the considered model that were not present in previously studied systems.
As can be seen in Fig.~\ref{fig:wtd1}(a), the transitions $\mathcal{W}_{\rm LL}$ where two electrons tunnel into the left FM lead
are strongly suppressed for short times $\tau \lesssim 1$
and have a maximum near $\tau \approx 1.5$.
This behavior is due to the fact that the two electrons are not able to instantaneously tunnel through the same junction.
On the other hand, the transitions to two different leads $\mathcal{W}_{\rm RL}$,
through the left and then through the right junction,
have a high finite value at short times, which approaches a maximum after passing time $\tau \approx 2$.
This finite WTD value at short times,
compared to the distribution of $\mathcal{W}_{\rm LL}$,
indicates the splitting processes present in the system,
as the tunneling through the left contact is quickly followed by the tunneling through the right one.
For longer times, $\tau \gtrsim 2$,
both $\mathcal{W}_{\rm LL}$ and $\mathcal{W}_{\rm RL}$ distributions start to
equilibrate and evolve slowly decaying in an irregular oscillatory manner.

\begin{figure}[t]
	\centering
	\includegraphics[width=\linewidth]{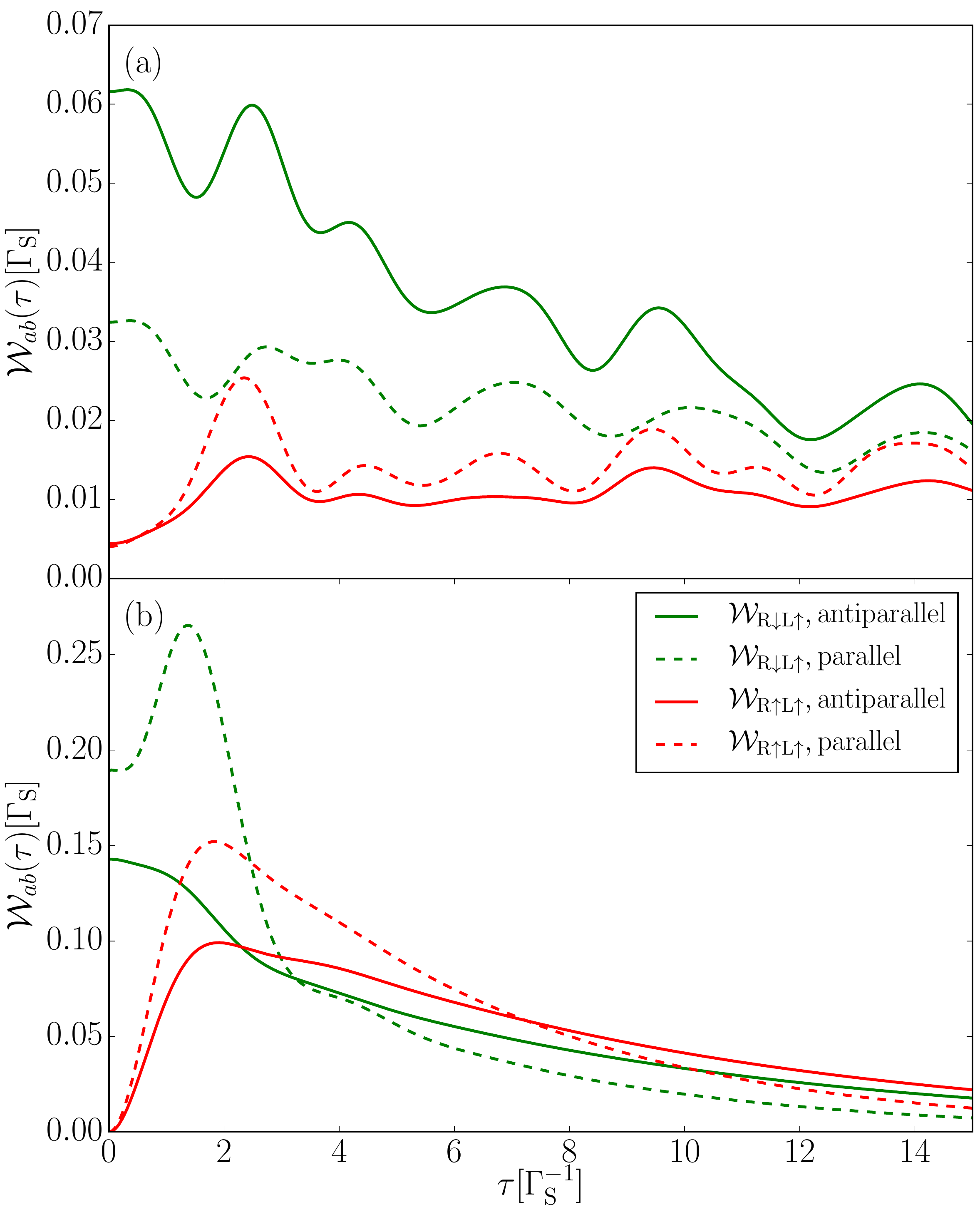}
	\caption{The spin-resolved electron waiting time distribution for
		the same parameters as in Fig. \ref{fig:wtd1}.
		Panel (a) shows the results when the coupling to ferromagnetic leads is equal to $\Gamma=\Gamma_{\rm S}/10$,
		while panel (b) is calculated for $\Gamma=10\Gamma_{\rm S}$.
		Dashed (solid) lines correspond to the parallel (antiparallel) configuration.}
	\label{fig:wtd2}
\end{figure}

If the ratio of couplings of TQD to FM leads and superconductor is inverted,
i.e. $\Gamma=10\Gamma_{\rm S}$,
the Cooper pairs are injected at slower rate than the rate of transferring electrons into the drain leads.
The WTD for this case is shown in Fig. \ref{fig:wtd1}(b)
for the same two types of transitions as shown in Fig. \ref{fig:wtd1}(a).
In consequence, the distribution in all the considered cases is quantitatively similar.
It exhibits a maximum for times proportional to the inverse
of injecting rate $\tau \approx  1/\Gamma_{\rm S}$
and quickly decays according to the coupling strength $\Gamma$.
The coherent oscillations between superconductor and TQD
or those due to internal dynamics of TQD subsystem are no longer exposed in the WTD.
Again, for transition $\mathcal{W}_{\rm LL}$ the distribution is completely
suppressed for $\tau \rightarrow 0$ and then rapidly evolves to a maximum.
On the other hand, the transition $\mathcal{W}_{\rm RL}$ has a significant finite probability
at short times with a maximum developing at earlier times compared to $\mathcal{W}_{\rm LL}$,
see Fig.~\ref{fig:wtd1}(b).
From these dependencies, one can expect to have a high splitting efficiency
at very short times $\tau \lesssim 1$, and that the splitting concerns a single Cooper pair,
as the consecutive Cooper pair is unlikely to be injected into the TQD system
before transferring both electrons into the drains.
Finally, for both considered coupling ratios,
in the case of antiparallel magnetic configuration
the distributions are enhanced at shorter times compared to the parallel one,
which is more optimal for efficient and fast Cooper pair splitting.

The insets in Fig.~\ref{fig:wtd1} present the waiting time distributions plotted in a logarithmic scale.
A rough approximation of the long-time behavior is performed
in order to designate the dependence on the coupling parameters.
The dependencies are $\mathcal{W}_{ab}(\tau)\sim e^{-A\tau\Gamma}$
and $\mathcal{W}_{ab}(\tau)\sim e^{-B\tau\Gamma_{\rm S}^2/\Gamma}$ for
the considered coupling ratios, $\Gamma=\Gamma_{\rm S}/10$ and $\Gamma=10\Gamma_{\rm S}$, respectively,
see Figs.~\ref{fig:wtd1}(a) and (b).
The parameters $A$ and $B$ are of the order of one and
include a non-trivial dependence on the system's magnetic configuration
and the internal parameters of the TQD.
We note that, quantitatively, the above predictions for the asymptotic behavior
of waiting time distributions are similar to the results obtained recently for the double quantum-dot-based
Cooper pair splitters.\cite{Walldorf2018}

In order to get a deeper understanding of the splitting processes,
we also determine the spin-resolved waiting time distributions.
Specifically, we analyze the distributions for transitions
$\mathcal{W}_{\rm R \downarrow L \uparrow }$ and $\mathcal{W}_{\rm R \uparrow L \uparrow }$,
where the former one corresponds to the scenario in which an electron with
spin $\sigma=\uparrow$ tunneled through to the left lead and then after time $\tau$
the electron of opposite spin $\sigma'=\downarrow$ tunneled through the right junction strongly indicating CAR processes,
while the latter one describes the situation when through the left and right junctions two electrons of spin $\sigma=\sigma'=\uparrow$ were transferred.
A high probability of $\mathcal{W}_{\rm R \uparrow L \uparrow }$ transitions at short times
is an unwanted characteristic, as it directly indicates that both electrons have to be from two different Cooper pairs.
The corresponding WTDs are presented in Fig. \ref{fig:wtd2}.

When the coupling to normal contacts is smaller than that to superconductor, see Fig.~\ref{fig:wtd2}(a) for $\Gamma=\Gamma_{\rm S}/10$,
the distribution shows that the antiparallel configuration is clearly more favorable for the splitting efficiency,
as split electrons are mainly of the opposite spins,
especially for short times where the maximum is present, up to times $\tau \approx 10$.
For longer times, all distributions are approaching comparable values.
When the system is in the parallel configuration, the opposite ($\downarrow \uparrow$) and the same ($\uparrow \uparrow$)
spin contributions of $\mathcal{W}_{\rm RL}$ are getting closer together,
but still clearly at short times $\mathcal{W}_{\rm R \downarrow L \uparrow }$ is dominant.

The characteristics are however less promising when the coupling to the FM leads
exceeds the coupling to superconductor, see Fig.~\ref{fig:wtd2}(b) for $\Gamma=10\Gamma_{\rm S}$.
For short times, when $\tau \lesssim 1$, the transitions $\mathcal{W}_{\rm R \downarrow L \uparrow }$ are dominating,
but generally the probability is higher in the parallel configuration compared to the antiparallel one.
On the other hand, the distribution in the antiparallel configuration, due to lower probabilities at shorter times,
is more spread and significant in a wider time range. $\mathcal{W}_{\rm R \uparrow L \uparrow }$ in this regime
is also strongly suppressed for short times $\tau \lesssim 1$,
although the distribution quickly raises to a maximum at $\tau \approx 2$ in both magnetic configurations.
Nonetheless, the maximum in the parallel configuration is higher
and results from a faster decay as the times elapses.
We recall the fact that the transition $\mathcal{W}_{\rm R \uparrow L \uparrow }$
requires two distinct Cooper pairs injected into the TQD subsystem.
Therefore, this transition is strongly suppressed for short times, $\tau \lesssim 1$,
as it is very unlikely that two Cooper pairs are provided in such short time,
while fast decay after the maximum is due to considerable value of the coupling to FM leads.

\subsubsection{Antisymmetric gate detuning ($\varepsilon_{\rm L}=-\varepsilon_{\rm R}=10\Gamma_{\rm S}$)}

The study of the Andreev current and its cross-correlation dependencies
for the antisymmetric detuning scheme has already shown that this configuration
is favorable for the efficient splitting of Cooper pairs.
The  analysis of WTDs presented in the sequel is generally in agreement with earlier conclusions,
especially when compared with the statistics in the case of symmetric setup discussed in previous section.

\begin{figure}[t]
\centering
\includegraphics[width=\linewidth]{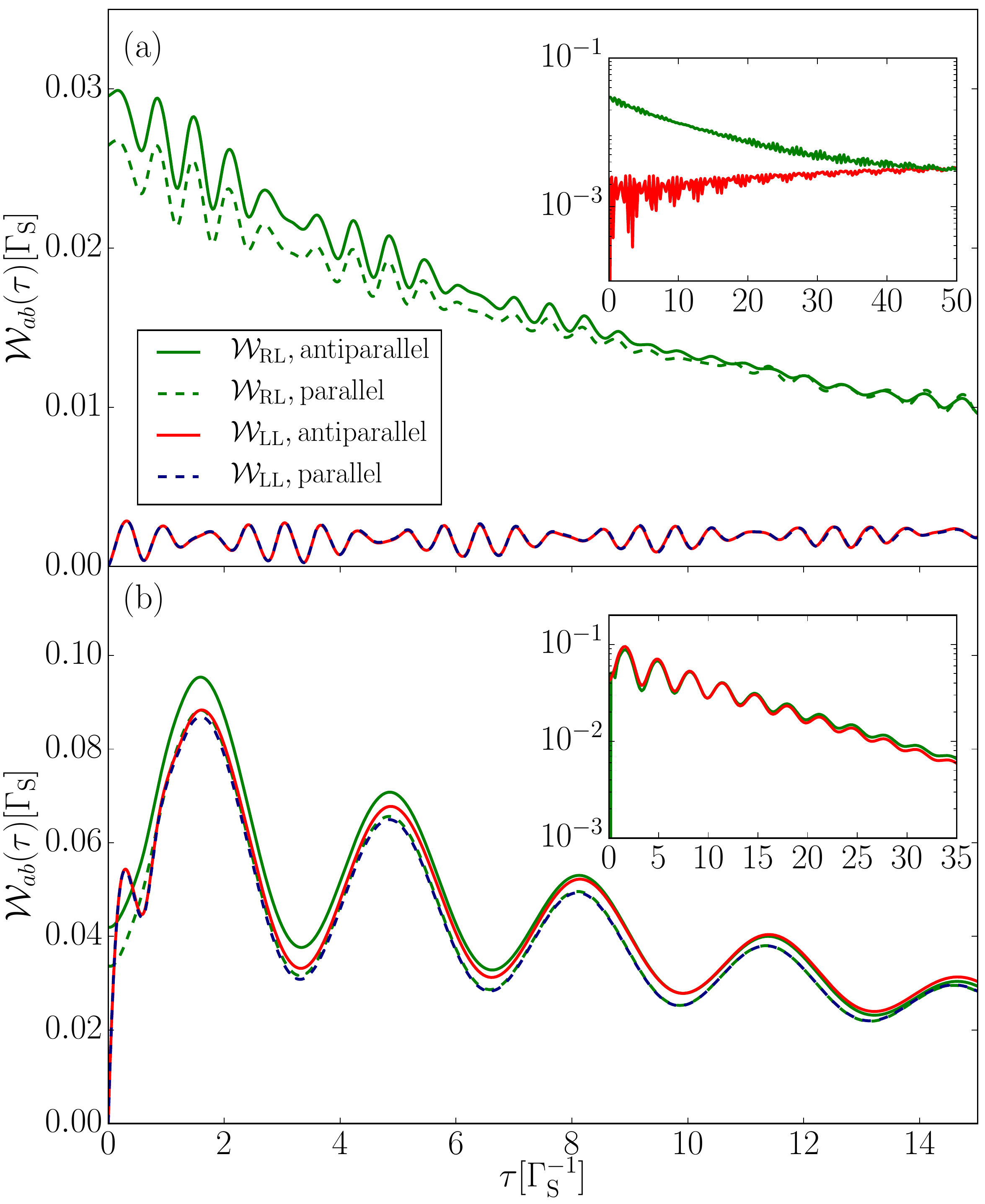}
\caption{The charge-resolved electron waiting time distribution
	for antisymmetric detuning of quantum dot energy levels.
	The parameters are: $\Gamma_{\rm S}\equiv 1$, $t=1$, $\varepsilon_{\rm M}=0$ and $\varepsilon_{\rm L}=-\varepsilon_{\rm R}=10\Gamma_{\rm S}$,
	in (a) $\Gamma=\Gamma_{\rm S}/10$, whereas in (b) $\Gamma=10 \Gamma_{\rm S}$.
The insets show the long-time asymptotic behavior of the WTDs.
The solid (dashed) lines correspond to the antiparallel (parallel) configuration.
}
\label{fig:wtd3}
\end{figure}

\begin{figure}[t]
	\centering
	\includegraphics[width=\linewidth]{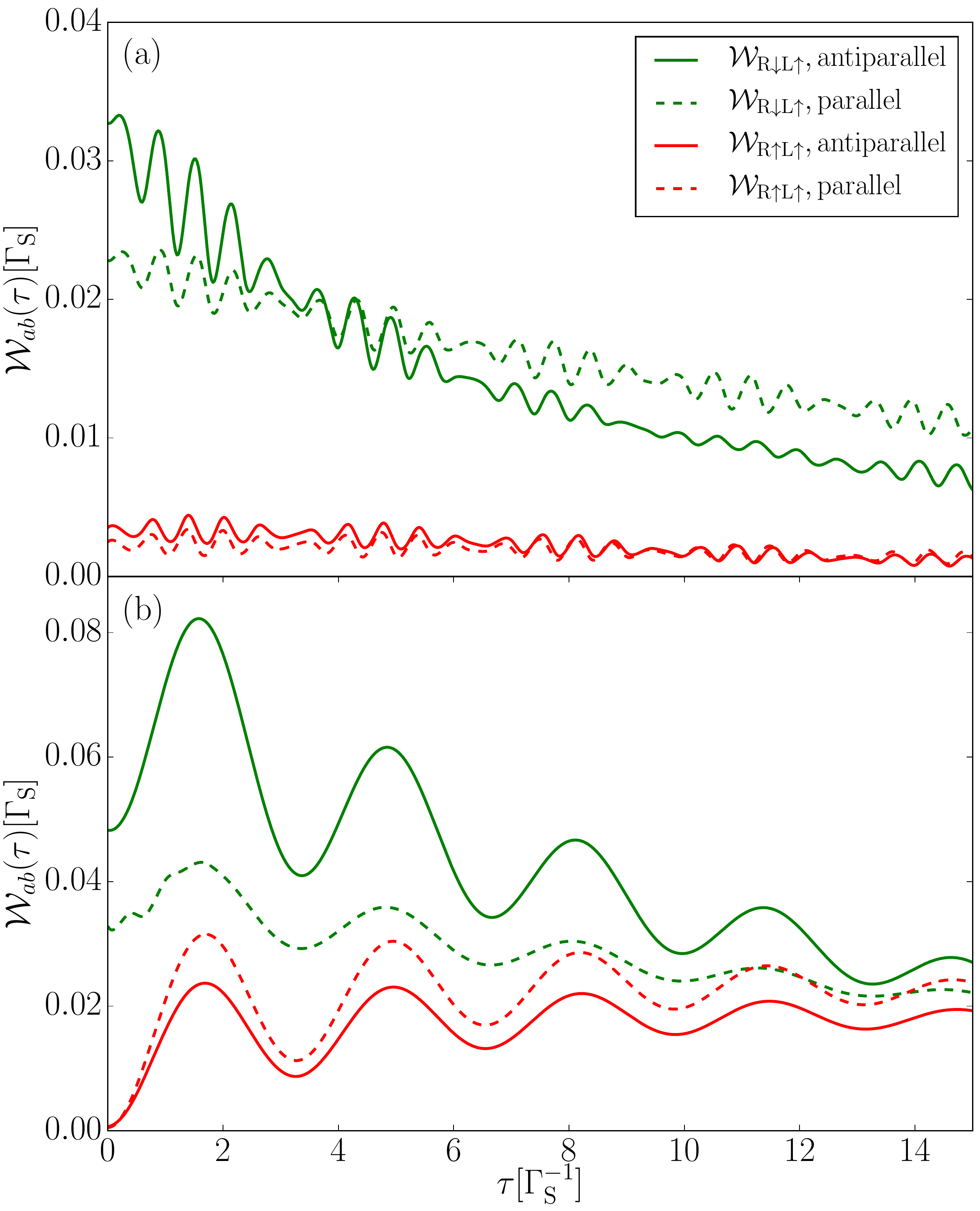}
	\caption{The same as in Fig. \ref{fig:wtd3} determined for spin-resolved electron waiting time distributions
		$\mathcal{W}_{\rm R \downarrow L \uparrow }$ and $\mathcal{W}_{\rm R \uparrow L \uparrow }$.}
	\label{fig:wtd4}
\end{figure}

In Fig.~\ref{fig:wtd3} the charge-resolved WTDs are shown for a strong antisymmetric detuning
$\varepsilon_{\rm L}=-\varepsilon_{\rm R}=10\Gamma_{\rm S}$. When the coupling ratio is in favor of the superconductor,
$\Gamma=\Gamma_{\rm S}/10$, fast coherent oscillations are again present in the distributions, see Fig.~\ref{fig:wtd3}(a).
Interestingly, due to the splitting of the Andreev levels by means of antisymmetric detuning,
additional oscillation frequency is present,
which results in the beats-like behavior, see the inset of Fig.~\ref{fig:wtd3}(a) with a logarithmic y-scale.
The distribution for splitting transitions $\mathcal{W}_{\rm RL}$
is strongly shifted towards short times, indicating fast emission of Cooper pairs through CAR processes.
Moreover, the transitions at short times are increased even further in the antiparallel configuration.
On the other hand, the waiting time distribution for DAR processes, $\mathcal{W}_{\rm LL}$, is characterized by a weakly oscillating flat distribution,
which is much smaller compared to $\mathcal{W}_{\rm RL}$ and is similar in both magnetic configurations.

For the reversed coupling situation, i.e. $\Gamma=10 \Gamma_{\rm S}$,
all distributions for the considered transitions are almost identical,  see Fig.~\ref{fig:wtd3}(b).
The significant difference is again at very short times, $\tau \lesssim 1$,
where the splitting transitions have a considerable value, while direct processes are suppressed.
The oscillations induced by the Andreev levels splitting
are evident in the considered case, although additional oscillations are not present
due to fast transfer of the electrons through the device.

The corresponding spin-resolved distributions are displayed in Fig.~\ref{fig:wtd4}.
For $\Gamma=\Gamma_{\rm S}/10$, see Fig.~\ref{fig:wtd4}(a),
the splitting transitions of opposite spins
$\mathcal{W}_{\rm R \downarrow L \uparrow }$ are strongly enhanced in the antiparallel configuration
for times $\tau \lesssim 4$, as compared to the parallel alignment.
The distribution of splitting transitions of electrons with the same spin
$\mathcal{W}_{\rm R \uparrow L \uparrow }$ remains intact
and is flat in the whole time range.
Similarly, when the coupling strengths are tuned such that
$\Gamma=10 \Gamma_{\rm S}$,
the spin-resolved WTDs expose significant increase of the splitting transitions
when the system is in the antiparallel configuration.

\section{Conclusions}\label{Conclusions}

In this paper we have analyzed the subgap transport properties of the triple quantum-dot-based Cooper pair splitters
with ferromagnetic contacts.
The Andreev current, associated differential conductance, tunnel magnetoresistance,
as well as current cross-correlations were calculated
by means of the real-time diagrammatic technique. Additionally, the electron waiting time distributions were found with the aid
of Markovian quantum master equation.
The main focus was set on the transport regime
where the device works optimally as a Cooper pair beam splitter, i.e.
when the Cooper pairs injected from the superconductor
are transferred into two separate ferromagnetic leads.
In particular, two different schemes of detuning the energy levels
of the dots placed in the arms of the splitter were considered:
the symmetric and antisymmetric one.
The former scheme exposes rather moderate Andreev transport properties, with one narrow regime
where the Andreev current and positive cross-correlations are considerable.
On the other hand, when the dots' levels are detuned antisymmetrically,
the device exhibits much more promising transport characteristics
as far as Cooper pair splitting is concerned,
with a possibility of maximizing the Andreev current
and obtaining high positive cross-correlations in a wide range of parameters.
The influence of inter-dot tunnel amplitude was also analyzed with conclusion that
considerable values of this parameter may lead to reduction of the device splitting efficiency in the low-bias regime.
Furthermore, in the antisymmetric detuning scheme,
a significantly better waiting time distributions for the splitting processes are predicted,
especially in antiparallel configuration of the leads.
It is important to emphasize that the efficiency enhancement of the CPS device due to
the use of ferromagnetic contacts revealed in WTD, is predicted in particular for short times.
We also note that in both detuning scenarios we found
the transport regime where the Andreev current is suppressed due
to a triplet blockade. In addition, positive tunnel magnetoresistance
is predicted in a wide range of dot level detunings and bias voltages.

We believe that the presented theoretical study extends the understanding of hybrid devices working
as quantum dot-based Cooper pair splitters, and gives a valuable guidance to tuning and optimizing experimental setups.
Moreover, the underlined importance of various correlations and interference effects on transport processes of the analyzed system
is expected to stimulate further research in this area.

\section{Acknowledgements}
We acknowledge useful discussions with Kacper Bocian.
This work was supported by the National Science Centre Poland from funds
awarded through the decisions No. DEC-2018/29/N/ST3/01038 (KW)
and DEC-2017/27/B/ST3/00621 (IW).

\section*{Appendix: Details of first-order diagrammatic calculations} \label{appendix}

Below, we describe the details
concerning calculations conducted with the aid of the real-time diagrammatic technique approach.
An important step is to build the matrix \textbf{W}
with matrix elements being the transition rates $W_{\chi' \chi}$.
For that, it is necessary to determine the corresponding self-energies.
In order to find the self-energies $\Sigma_{\chi \chi'}$,
one needs to evaluate all irreducible, topologically different diagrams describing the tunneling processes.
The contribution of each diagram is calculated by applying the diagrammatic rules
\cite{Schoeller1994, Konig1996}.
For the description of the sequential tunneling transport regime,
it is necessary to calculate the self-energies within the first-order
perturbation expansion with respect to the tunnel coupling $\Gamma$ to ferromagnetic leads.
The first-order diagrams contain a single tunneling line.
Here we present an exemplary diagram contributing to the
self-energy $\Sigma^{(1)}_{\chi(N) \chi'(N+1)}$.
\begin{eqnarray*}
\includegraphics[width=.4\columnwidth,left]{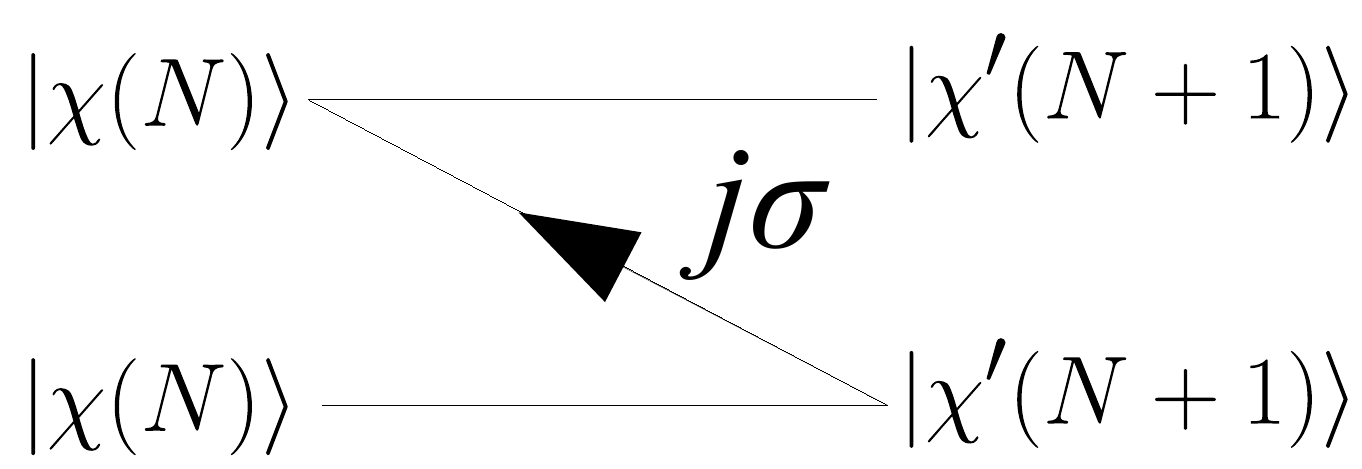}\nonumber\\
= (-1)^1 
\int \! d \omega \frac{\gamma_{j \sigma}(\omega)}{\omega - \varepsilon_{\chi'} + \varepsilon_{\chi}+i\eta}
| \langle \chi' | d^{\dagger}_{j \sigma} | \chi \rangle |^2,
\end{eqnarray*}
where $\gamma_{j \sigma}(\omega)=\frac{\Gamma^{\sigma}_{j}}{2\pi}f_{j}(\omega)$
is a factor associated with tunneling line, while
$f_{j}(\omega)$ is the Fermi-Dirac distribution of lead $j$ and $\eta = 0^{+}$.
The above diagram accounts for an electron with spin $\sigma$
tunneling from the lead $j$, between states $|\chi(N)\rangle$ and $|\chi'(N+1)\rangle$.
$N$ is the total occupation number of TQD subsystem defined as $N=\sum_{j\sigma}n_{j\sigma}$.
All topologically different diagrams need to be evaluated
and their contributions summed according to the $|\chi\rangle$ and $|\chi'\rangle$ states.
Then, the self-energy $\Sigma_{\chi(N) \chi'(N+1)}$ is given by

\begin{equation*}
\Sigma_{\chi(N) \chi'(N+1)} = 2\pi i \sum_{j \sigma}
\gamma_{j \sigma}(\varepsilon_{\chi'} - \varepsilon_\chi)
| \langle \chi' | d^{\dagger}_{j \sigma} | \chi \rangle |^2.
\end{equation*}
The other self-energies can be calculated in a similar fashion.
\cite{Schoeller1994, Konig1996}.
Then, the self-energy is related to the elements of matrix $\mathbf{W}$ through
$\Sigma_{\chi \chi'}= iW_{\chi' \chi}$.

Finally, the matrix elements of $\mathbf{W}^{I_j}$, i.e. the self-energy matrix
contributing to the current, are given by
\begin{equation}
\begin{aligned}
W^{I_j}_{\chi(N) \chi'(N+1)} &= - \sum_{\sigma}
[1-f_j(\varepsilon_{\chi'} - \varepsilon_\chi)]\Gamma_j^{\sigma}| \langle \chi | d_{j \sigma} | \chi' \rangle |^2,\\
W^{I_j}_{\chi(N) \chi'(N-1)} &=  \sum_{\sigma}
f_j(\varepsilon_{\chi} - \varepsilon_\chi')\Gamma_j^{\sigma}| \langle \chi | d^\dagger_{j \sigma} | \chi' \rangle |^2,
\end{aligned}
\end{equation}
with $W_{\chi \chi}^{I_j}=0$.

\end{document}